\documentclass{aa}  

\usepackage{graphicx}
\usepackage{float}
\usepackage{txfonts}
\usepackage{comment}
\usepackage{xcolor}
\usepackage{tikz}

\defcitealias{2023ApJ...942L..13L}{LR23}
\usepackage[breaklinks=true]{hyperref}
\begin{document} 
   \title{A multi-scale view of molecular gas and magnetic fields in the Antennae galaxies}

   \subtitle{}

   \author{Akhil Lasrado\inst{1}
          \and
          Enrique Lopez-Rodriguez\inst{2}
          \and
          Claudia Cicone\inst{1}
          \and
          Axel Weiss\inst{3}
          \and
          Andreas Lundgren\inst{4}
          \and
          Boy Lankhaar\inst{1}
          \and
          Robert Wissing\inst{1}
          \and
          Sijing Shen\inst{1}}

   \institute{\inst{1}Institute of Theoretical Astrophysics, University          of Oslo, P.O. Box 1029, Blindern, 0315 Oslo, Norway\\
            \inst{2} Department of Physics and Astronomy, University of South Carolina, Columbia, SC 29208, USA\\
            \inst{3} Max-Planck-Institut für Radioastronomie, Auf dem Hügel 69 D-53121 Bonn, Germany \\
            \inst{4} Aix Marseille Univ, CNRS, CNES, LAM, Marseille, France\\
            \email{akhil.lasrado@astro.uio.no} \\}

   \date{}

 
  \abstract {Major mergers are transformative events in galaxy evolution, dynamically reshaping the stellar and gas makeup of the hosts, as well as the galaxy-scale magnetic fields (B-fields) that permeate their interstellar medium (ISM), altering both B-field strength and configuration. In this work, we study in detail the link between the co-evolving gas and B-fields in the cold ISM phase of the nearby Antennae galaxy merger (NGC 4038/4039), using CO(3--2) data from the LAsMA receiver at the APEX telescope and ALMA, together with archival 154 $\mu$m dust polarization data from the SOFIA/HAWC+. We characterize trends between measurements of molecular gas intensity and velocity dispersion with polarization measurements at kpc-scales with the APEX and SOFIA data. We find anti-correlations with CO intensity ($\rho\lesssim-0.6$) and CO velocity dispersion ($\rho\lesssim-0.4$), which are signposts of geometric depolarization. Since dust polarization measurements are a density-weighted line-of-sight integrated average of B-field vectors below the beam size, the ALMA data can serve as a proxy of density structures below the SOFIA beam on a channel-by-channel basis. Through a geometric analysis of structures in the ALMA channels, we identify salient kinematic features in the overlap region strongly aligned with the SOFIA-inferred B-field orientation. This indicates that these kinematic features might dominantly lend their orientation to the measured field direction, assuming a correspondence between molecular gas structures and galactic B-fields, owing to their density or high polarization fraction. We also identify a modest trend ($\rho\sim0.5$) between the coherence of the molecular gas, a proxy of the line-of-sight angular dispersion, at small scales, and the SOFIA polarization fraction, showing that the polarization fraction measured at $\sim2$ kpc scales is sensitive to structures at $\sim50$ pc scales.}


   \maketitle 
%

\section{Introduction}

Magnetic fields are dynamically important in galaxies. Polarimetry observations of nearby galaxies show them harboring magnetic (B-)fields with measured strengths of several $\sim\mu$G in the diffuse gas of normal galaxies \citep{2004A&A...417..541C, 2010ASPC..438..197F, 2013A&A...555A..23A, 2015A&ARv..24....4B} going to upto mG strengths measured in the molecular gas in starbursts such (eg. Arp 220, M82) \citep{2006ApJ...645..186T, 2008ApJ...680..981R}, being in close equipartition with the thermal and turbulent energies in the interstellar medium (ISM) \citep{2009ApJ...702.1230T, 2015A&A...578A..93B}. B-fields can strongly influence galaxy evolution, regulating galactic star formation \citep{2007ARA&A..45..565M, 2019FrASS...6....7K}, affecting cosmic ray transport \citep{2012MNRAS.423.2374U}, and guiding gas inflows \citep{2005A&A...444..739B, 2024A&A...691A.303T} and galaxy-scale outflows \citep{2026ApJ...996L..27L}.

A natural question that arises concerns the amplification of cosmological B-fields from weak primordial strengths of $\sim10^{-20}$ G in the early Universe, generated via the Biermann battery mechanism or plasma instabilities \citep{1950ZNatA...5...65B, 1959PhRvL...2...83W}, to their present day $\mu$G strengths. A commonly invoked remedy are turbulent dynamo mechanisms which convert turbulent kinetic energy into magnetic energy \citep{1998MNRAS.294..718S, 2023ARA&A..61..561B}. Turbulent dynamos are broadly categorized into two types, based on the scales on which they amplify B-fields relative to the turbulence injection scale: (i) small-scale (or fluctuation) dynamos that operate through the stretching, twisting, and folding of field lines, and which amplify fields on e-folding times of 10-100 Myr \citep{1972SvPhU..15..159V, 2004ApJ...612..276S, 2017MNRAS.469.3185P}, and (ii) large-scale (or mean-field) dynamos that, in the presence of large-scale asymmetries in the density and kinematics of the gas, can amplify B-fields on the order of a Gyr \citep{1997ApJ...475..263V, 2008A&A...490..501J, 2015PhRvL.115q5003S}.

Major galaxy mergers provide the necessary ingredients for efficient turbulent dynamo activity, and are therefore considered crucial for B-field amplification \citep{2011A&A...533A..22D, 2012MNRAS.419.3571G, 2014A&A...566A..40M, 2016A&A...593A..89R, 2018MNRAS.479.3343M}. Alongside altering B-fields, mergers greatly reconfigure the gas of the host galaxies, relieving it of its angular momentum which leads to galaxy-scale gas inflows, shocks, and tidal structures \citep{1987AJ.....93.1011K, 1992ARA&A..30..705B, 1994A&A...281..725C, 1996ApJ...464..641M}. The B-fields themselves couple to these gas flows, although the extent of this coupling is expected to vary significantly depending on the magnetic diffusion coefficient and the extent of turbulence injected in the ISM \citep{2001A&A...378...40S,2016A&A...585A..21F,2025JATIS..11c1620M}. Since galaxy mergers possess strong gas flows and B-fields, they offer a dynamic environment in which observations could be incisive in deciphering the degree to which ISM B-fields respond to strong gas flows.

Studies of the plane-of-sky orientation of B-fields in nearby galaxies have been largely enabled by radio synchrotron observations, which are known to trace B-fields in the volume-filling, warm and diffuse ISM \citep{2013pss5.book..641B, 2019Galax...8....4B}. On the other hand, (far-)infrared (FIR) dust polarization observations have been widely used to probe B-fields much closer to the cold and dense ISM. This is due to the alignment of dust grains by radiative torques (RATs) wherein elongated, non-spherical dust grains are preferentially aligned with their short axis parallel to the local magnetic field direction in the presence of an anisotropic radiation field. The dust then reprocesses UV radiation and emits polarized light in the infrared \citep{2015ARA&A..53..501A, 2016ApJ...831..159H}. This method has been used to study B-fields in the Milky Way and in nearby galaxies \citep{1964LowOB...6..123E,2018ApJ...861L..23L,2019ApJ...870L...9J,2022ApJ...936...92L}. It is important to note here that since these radio/FIR observations trace continuum polarization, the measured B-field orientation is a line-of-sight integrated measurement, which in the case of complex gas distributions, such as in galaxy mergers, can make the inferred polarization -- and thus magnetic field -- direction non-trivial to interpret \citep{2025ApJ...991..156P}.

In this work we characterize the molecular gas -- magnetic field interplay using new observations of the Antennae galaxies (NGC 4038/4039): a nearby ($D_{L} =22$ Mpc, \cite{2008AJ....136.1482S}) ongoing major merger undergoing an intense starburst (SFR $>10$ M$_{\odot}$ yr$^{-1}$, \cite{2010MNRAS.401.1839Z}) in an interaction region bridging the galaxies, termed the overlap region \citep{1990ApJ...349..492S}. The overlap region is the brightest region in CO emission \cite{2003ApJ...588..243Z, 2012ApJ...745...65U} and hosts molecular cloud complexes with high surface densities and pressures compared to normal galaxies \citep{2024ApJ...964..166K}, indicative of a gas-rich environment strongly affected by turbulence \citep{2014MNRAS.442L..33R}. Such conditions are conducive for magnetic field amplification. Indeed, \cite{2004A&A...417..541C} in a study of synchrotron polarization in this source estimated strengths of up to 30 $\mu$G, $\sim$3 times larger than that of regular spirals. In a more recent study, \cite{2023ApJ...942L..13L} (henceforth: LR23) use dust polarization measurements in addition to radio polarization data and find evidence of a turbulent dynamo operating in the overlap region. Such a region -- where strong B-fields likely affect gas flows and vice versa -- thus represents ideal conditions in which to study the interplay of the two phenomena.


In order to characterize this interplay, our work utilizes observations of the Antennae galaxies at multiple resolutions. We use new APEX/LAsMA observations of the CO(3--2) line together with SOFIA/HAWC+ data of the 154 $\mu$m far-infrared (FIR) continuum polarization from LR23 to infer variations of gas properties with polarization measurements across the galaxy at a $\sim2$ kpc scale. We also use archival high resolution ($\sim$$50$ pc) ALMA CO(3--2) observations from \cite{2024ApJ...971..176H} and explore how the geometry of molecular gas structures at small scales might manifest in the low resolution magnetic field measurements seen by SOFIA.

We structure the paper as follows: In Sect. \ref{sec:obs} we describe the observations used in the work. In Sect. \ref{sec:res} we present results of the APEX-SOFIA comparison and of geometrical comparisons of the ALMA and SOFIA data, the outcomes of which we discuss in Sect. \ref{sec:disc}. Our conclusions are listed in Sect. \ref{sec:conclusion}. We outline our methodology for comparisons of the datasets in Appendix \ref{sec:methods}.



\section{Observations}\label{sec:obs}
\subsection{APEX/LAsMA observations}\label{sec:obs_apex}
We observed the Antennae galaxies ($z=0.00547$) with the Atacama Pathfinder EXperiment (APEX), a 12-meter single dish sub-millimeter telescope located in Chile \citep{2006A&A...454L..13G} between 8 August and 20 August 2023, with an on-source observing time of 4.1 hours in median weather conditions (precipitable water vapor content $\sim$1.14 mm). The galaxy was mapped in the CO(3--2) emission line ($\nu_{\text{obs}} = 343.90$ GHz) with the 7-pixel heterodyne array receiver LAsMA (large APEX sub-millimeter array\footnote{\href{https://www.mpifr-bonn.mpg.de/5278286/lasma}{https://www.mpifr-bonn.mpg.de/5278286/lasma}}), which enables a faster mapping speed than traditional single pixel receivers, in total power OTF (on-the-fly) mode in a rectangular pattern covering a large field-of-view of $7^{\prime}\times7^{\prime}$ centered on the galaxy ($\alpha_{\rm 2000} = 12^{\rm h}01^{\rm m}53^{\rm s}$, $\delta_{\rm J2000} =  -18^{\circ}52'38''$). All quoted velocities are redshift corrected. Data reduction was performed using \textsc{GILDAS/CLASS}\footnote{\href{http://www.iram.fr/IRAMFR/GILDAS}{http://www.iram.fr/IRAMFR/GILDAS}} followed by further masking in \texttt{python}.

On an initial inspection of the spectra we noticed line absorption at several positions in the field at $\nu_{\text{obs}}\sim$343.2 GHz, corresponding to $\rm{v}\in(+256,+705)$ km/s, just outside the range of observed CO line emission. We identified these as atmospheric ozone lines and flagged them in \texttt{CLASS} in each scan. A zeroth-order baseline was also subtracted from each scan, by masking  the CO(3--2) line from $v\in(-300,300)$ km s$^{-1}$. For further reduction, we selected a bandwidth of 2.6 GHz (3000 km/s) centered on the rest frequency of CO(3--2), masked the line emission in a window in the range $v \in (-500, 500)$ km/s, and subtracted a first order baseline from the spectra (in $T_{A}^{*}$) for each scan. The spectra were then re-binned from a native spectral resolution of 0.43 km/s to a resolution of 5 km/s. Before constructing the final cube, we also flagged scans from the receiver backend APL-309-F402 (retaining APL-308-F402 to APL-313-F402) due to baseline instabilities (comprising $\sim$14\% of all scans). To convert the cube to the main beam brightness temperature scale, we used a main beam efficiency of $\eta_{\text{MB}} = 0.77\pm0.06$ for August 2023\footnote{\href{www.apextelescope.org/telescope/efficiency/}{www.apextelescope.org/telescope/efficiency/}}, and use a Kelvin to Jansky conversion factor of $37\pm3$ Jy/K for conversion to physical units. The final cube has a median rms value of $\sim$12 mK in $T_{A}^{*}$ units ($\sim$0.44 Jy/beam), and has a beam FWHM of 19.1$^{\prime\prime}$ ($\sim$2 kpc at a distance of 22 Mpc for the Antennae).

In order to extract low surface brightness emission and to construct robust moment maps from the APEX and ALMA CO(3--2) data cubes, we follow the dilated masking procedure employed in \cite{2024A&A...682A.137G}, and refer to the latter for a detailed description of the method. Briefly, the method first selects voxels in the cube above a high signal-to-noise (S/N) threshold ($R_{1}$) after which connected voxels down to a second, lower S/N threshold ($R_{2}$) are included. A third S/N cut ($R_{3}$) along the velocity axis is also applied. This ensures that low S/N voxels included in the mask are still associated with high S/N regions. We then calculate moments using the masked data cubes. We obtain the velocity dispersion map as the square root of the second moment map. For the APEX data, we select thresholding values of \{$R_{1}$, $R_{2}$, $R_{3}$\} = \{3, 2, 3\}. The resulting moment maps are shown in Fig. \ref{fig:presentation}.


\subsection{SOFIA/HAWC+ observations}\label{sec:obs_sofia}

SOFIA (Stratospheric Observatory for Infrared Astronomy) observed the Antennae galaxies as part of the SALSA (Survey of Extragalactic Magnetism with SOFIA) survey (ID: 08\_0012; PIs: E. Lopez-Rodriguez and S. A. Mao), on 2 June 2022 and 11 June 2022 from Palmdale, USA and 21 June 2022 and 27 June 2022 from Christchurch, New Zealand. The observations were performed with the HAWC+ (High-resolution Wide-Band Camera Plus) at 154~$\mu$m with on-the-fly-mapping polarimetric mode with
a scan rate of 200$''$ s$^{-1}$, a scan amplitude of 100$''$ $\times$ 100$''$, and a scan on-source time of 120s per halfwave plate position angle.

The data was reduced using the pipeline scheme described in SALSA III \citep{2022ApJ...936...65L}, and contain the Stokes $IQU$ products, polarization fraction ($p$), position angle ($PA$), polarized intensity ($PI$), and their uncertainties, with the pixel scale and beam of the APEX data ($3''/19''$) (the native resolution of the SOFIA data is 13.6$''$ at 154 $\mu$m). The polarization measurements were then de-biased \citep{1962AdA&A...1..289S} and corrected for instrumental polarization. The final data products have an on-source time of 4.8 hours. The flux calibration uncertainty in the Stokes I map is estimated to be $8\%$, and an intrinsic polarization uncertainty of 0.2\%, and angular uncertainty of 3$^{\circ}$ are also included \citep{2018JAI.....740008H, 2022ApJ...936...65L}. The polarization fraction map (shown in Figure \ref{fig:presentation}) was computed using measurements with $I/\sigma_{I}>100$. Calculation of polarization measurements and their uncertainties in individual sight-lines are described in Sect. \ref{sec:methods_hexaprops}.

\subsection{Archival ALMA observations}

We supplement our analysis with archival data from the Atacama Large Millimeter/submillimeter array (ALMA) from \cite{2024ApJ...971..176H} of the CO(3--2) line. To recover the most extended features in the ALMA data and mitigate negative sidelobes, we used the \texttt{feather} task in \texttt{CASA} to combine visibilities from the TP, 7m, and 12m arrays, using a default single dish factor (\texttt{sdfactor}) of 1. We note here that while the latter factor is known to be uncertain \citep{2023PASP..135c4501P}, we do not use the combined data with the goal of deriving fluxes, and are therefore not strictly concerned with accuracy in flux recovery. To improve the signal-to-noise of the data, we additionally re-bin the data to spectral channels of widths of 10 km s$^{-1}$. The beam size of the data is 0.52$''$ (55 pc at 22 Mpc). We use the same dilated masking routine as used for the APEX data to extract moment maps from the ALMA data, with thresholding values of \{$R_{1}$, $R_{2}$, $R_{3}$\} = \{10, 4, 4\}.

\section{Results}\label{sec:res}

\subsection{Far-infrared and sub-millimeter maps of the Antennae merger}

\begin{figure*}[ht]
\centering
\includegraphics[width=1\textwidth]{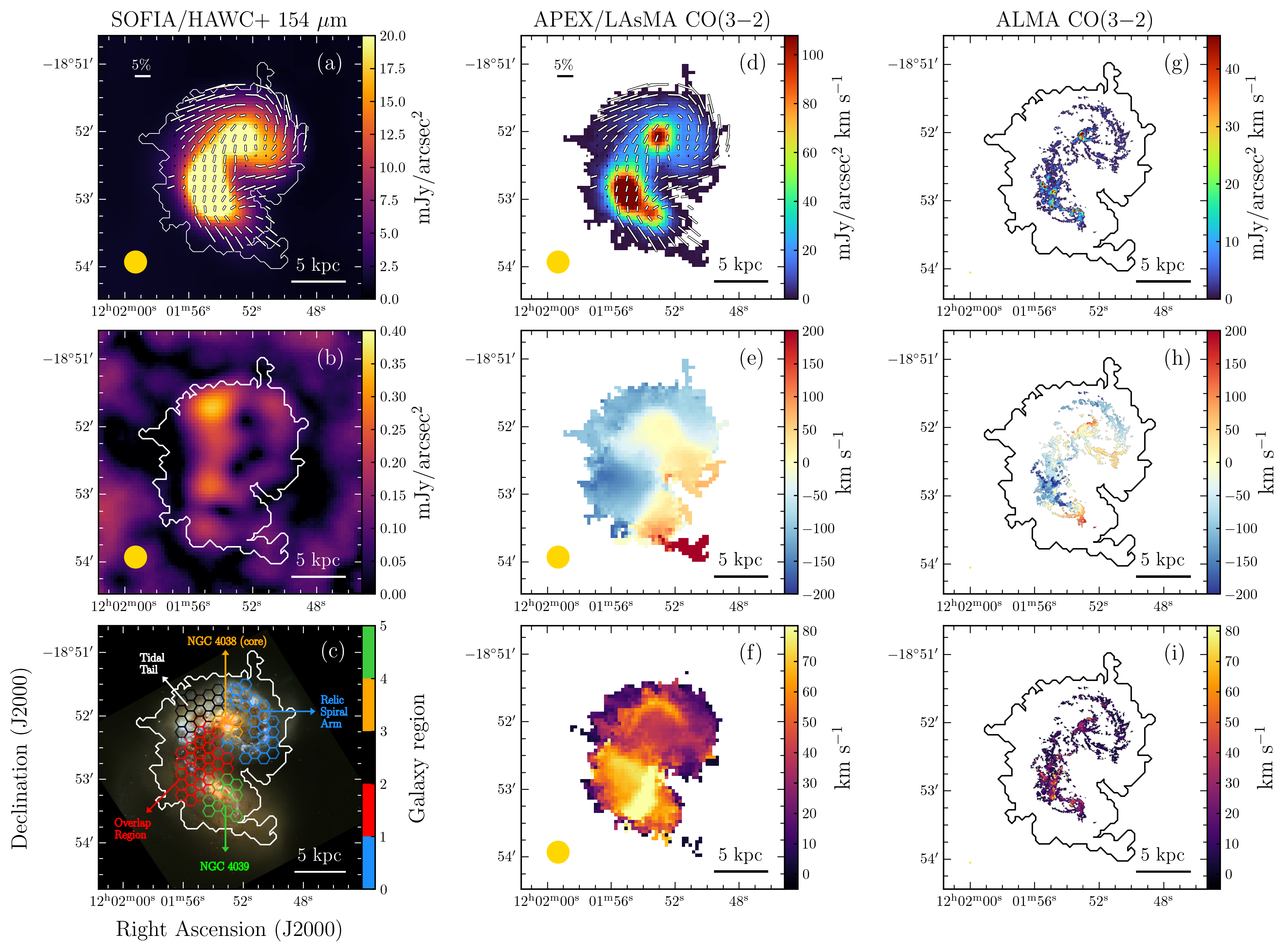}
\caption{Overview of data used in this work. 
The first column shows (a) the HAWC+ Stokes $I$ intensity, (b) the polarized intensity, and (c) an RGB Hubble image constructed from the 435 nm, 550 nm, and 814 nm WFPC2 filters from top to bottom, respectively. Each plot shows the outline of the APEX integrated intensity map. The Stokes $I$ map is overlaid with polarization vectors from HAWC+ rotated by 90$^{\circ}$ to show the local field orientation, scaled to the degree of polarization. The vectors have been rescaled to the square root of the polarization fraction for visualization purposes. The Hubble image is overlaid with the hexagonal apertures used in our analysis of SOFIA and APEX measurements, colored according to the region they are assigned to, including the tidal tails (black), the relic northern spiral arm (blue), the overlap region (red), the southern galaxy (NGC 4039) (green), and the northern galaxy (NGC 4038) core (orange). The second column shows (d) the integrated intensity, (e) first moment, and (f) velocity dispersion map derived from APEX.  The third column shows (g) the integrated intensity, (h) first moment, and (i) velocity dispersion map of the ALMA data.  Each plot includes an outline of the APEX integrated intensity map.}
\label{fig:presentation}
\end{figure*}

We present maps from the SOFIA, APEX, and ALMA data in Figure \ref{fig:presentation}. The first column shows measured B-field orientations on the Stokes I intensity map, and the polarized intensity maps from SOFIA. The vectors show the B-field following the spiral arm of the northern galaxy (NGC 4038), bridging the northern and southern galaxy (NGC 4039) both in and towards the west of the overlap region, and parallel to the major axis of NGC 4039 in the south. The polarization fraction is enhanced in the tidal tails and diminished in the overlap region. The third plot in the first column shows an optical image from Hubble of the galaxies, overlaid with a hexagonal grid that we use to compare properties across the data (\ref{sec:res_APEX_SOFIA_props}). The grid is split into salient regions of the Antennae, namely, the relic spiral arm of the northern galaxy, the tidal tail, the overlap region, the southern galaxy, and the core of the northern galaxy.

The moment maps derived from the APEX and ALMA data are presented in the second and third columns, respectively. The integrated intensity map of APEX peaks in the overlap region followed by the cores of the northern and southern galaxies. The enhanced CO intensity in the overlap region, as seen in the ALMA map, arises primarily from bright molecular clouds in the region, termed ``super GMCs'' \citep{2000ApJ...542..120W}. We note here that the goal of obtaining the large FoV ($7'\times7'$ or $45\times45$ kpc$^{2}$) map of the galaxies was to look for large-scale diffuse and extended emission, as in \cite{2025A&A...701A.235L}, such as a molecular gas counterpart of the prominent HI tidal tails. Our derived integrated intensity map (Fig. \ref{fig:presentation}, (\textit{d})), does not show prominent CO(3--2) emission in the tidal tails, or in other extended features.

In the APEX first moment map, we see large velocity variations in the southern galaxy and in the relic spiral arm. Towards the southern galaxy the velocity variation reflects the blue-shifted ($\sim$$-50$ km s$^{-1}$) nature of gas in the overlap region with respect to the southern galaxy ($\sim$$+25$ km s$^{-1}$). A similar large-scale velocity difference appears between the relic spiral arm and the rest of the northern galaxy. This apparent decoupling is interestingly in not following a typical rotational signature, despite the northern galaxy appearing to be less disturbed compared to the rest of the system. These trends are also observed in the ALMA first moment map.

Such velocity variations that occur across roughly one APEX beam manifest as high dispersion features in the APEX velocity dispersion map, most notably in the region between the overlap region and southern galaxy, and at the interface of the relic spiral arm and the rest of the northern galaxy. We verify that the southern high dispersion feature does not arise from potential outflows from the absence of high velocity tails/features in beam-averaged spectra along the feature, shown in Figure \ref{fig:highdisp_spec}. The high $\sigma_{\rm CO}$ measured in these regions therefore includes contributions from kinematically distinct regions, rather than intrinsic cloud velocity variations along the line-of-sight (LOS), which is resolved out in the ALMA velocity dispersion maps, where similar large values of $\sigma_{\rm CO} \gtrsim 70$ km s$^{-1}$ are not observed.

We can make preliminary inferences of the relations between the polarization and spectroscopic data from these maps. The total FIR and total CO intensities are clearly correlated, peaking in the overlap region. This agreement can be sourced from the Kennicutt-Schmidt relation \citep{1998ApJ...498..541K}, assuming the FIR intensity and CO intensity to be proxies of the star formation rate surface density and molecular gas surface density, respectively. The degree of polarization compared to the total CO intensity also appears to be anti-correlated to first order, since the large $p$ features in the tidal tail have low $I_{\rm CO}$, while the overlap region where $I_{\rm CO}$ peaks has very low $p$. We quantify these correlations in the following section.

\subsubsection{Molecular gas mass calculation and APEX/ALMA flux comparison}\label{sec:res_flux_calc}

We measure a total CO(3--2) flux from the APEX data of $10900\pm1200$ Jy km s$^{-1}$, equivalent to a CO(1--0) flux of $2400\pm550$ Jy km s$^{-1}$ using a CO(3--2)/CO(1--0) line ratio of $0.5\pm0.2$ \citep{2012ApJ...745...65U, 2024ApJ...971..176H}. The molecular gas mass obtained using a galactic CO-to-H$_{2}$ conversion factor $\alpha_{\rm CO}=4.35\pm1.3$ $M_{\odot}$ (K km s$^{-1}$ pc$^{2}$)$^{-1}$ is $(1.2\pm0.6)\times10^{10}$ $M_{\odot}$, which is consistent with previous measurements \citep{2001ApJ...548..172G, 2003ApJ...588..243Z}. We compute a upper limit on the CO(3--2) flux in the tidal tails by stacking spectra in a mask covering the tidal tails seen in HI using archival BIMA data, and assuming a linewidth of $\sim$10 km s$^{-1}$ in the tidal tails (similar to the linewidth of the HI, from \cite{2001AJ....122.2969H}), and obtain a 3$\sigma$ upper limit of $\sim$80 Jy km s$^{-1}$, which corresponds to an $L'_{CO}=10^{7}$ K km s$^{-1}$ pc$^{2}$. The ALMA data from the combination of the 12m + 7m + TP visibilities yields an integrated flux of $8800\pm900$ Jy km s$^{-1}$, which is $\sim80\%$ of the flux derived from APEX. 

\subsection{Comparison of APEX and SOFIA properties}\label{sec:res_APEX_SOFIA_props}

\begin{figure*}
\centering
\includegraphics[width=1\textwidth]{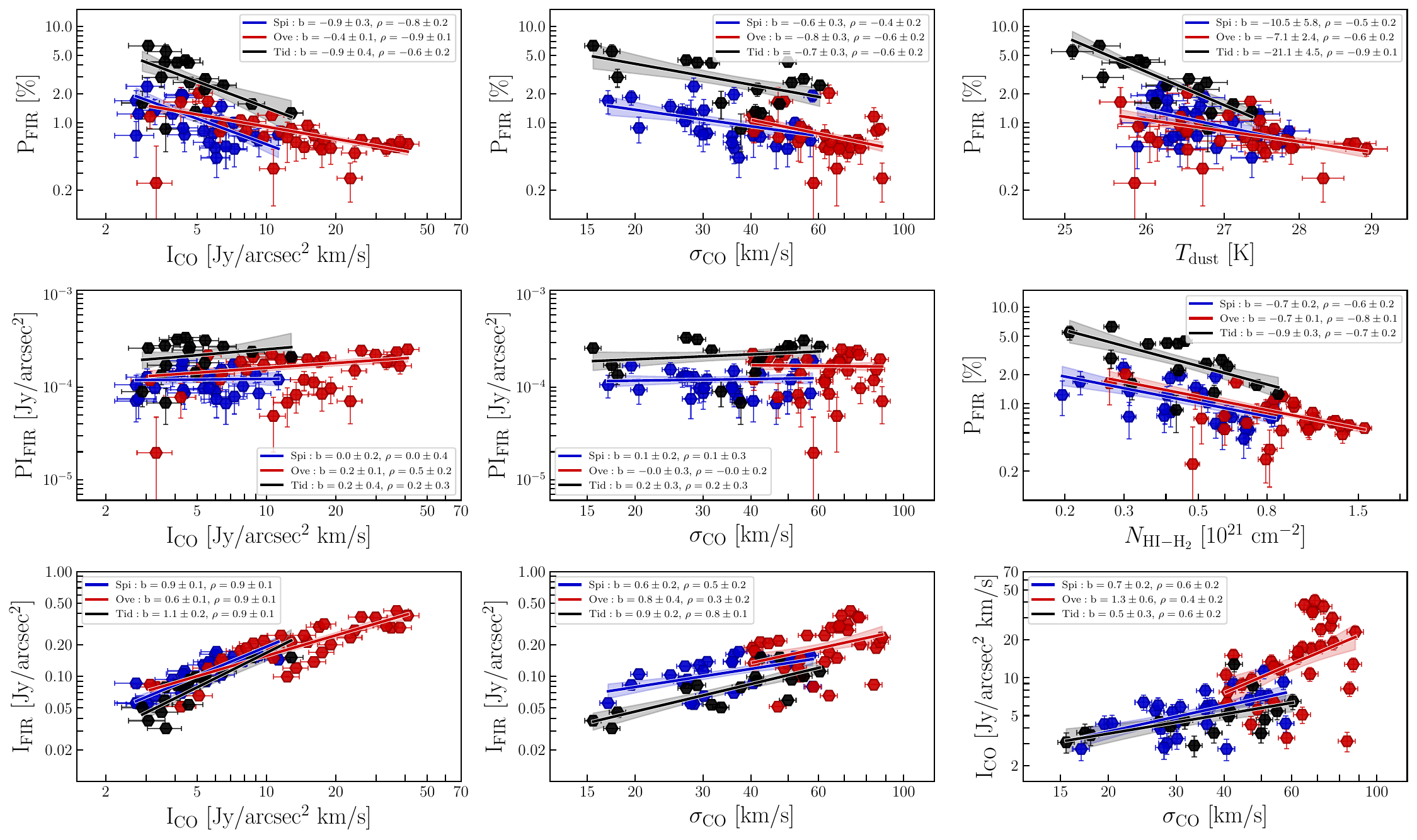}
\caption{Molecular gas measurements from APEX versus polarization quantities from SOFIA for the relic spiral arm, the overlap region, and the tidal tails. From top to bottom, the left column shows $p$, $PI$, $I_{\rm FIR}$ vs $I_{\rm CO}$, the middle column shows $p$, $PI$, $I_{\rm FIR}$ vs $\sigma_{\rm CO}$, and the right column shows $p$ vs $T_{\rm dust}$, $N_{\rm HI-H_{2}}$, and $I_{\rm CO}$ vs $\sigma_{\rm CO}$. Each data point represents a measurement performed from a single hexagonal aperture from the grid shown in Fig. \ref{fig:presentation}c. The color indicates the corresponding region in the galaxy merger: red for the overlap region, blue for the northern relic spiral arm, and black for the tidal tails. The relations are fit independently for each region using a linear fit, and the best-fit relations are plotted with their 1$\sigma$ uncertainty. The derived slope and correlation coefficient for each fit are reported in the legends and are calculated accounting for errors in both variables using the \texttt{linmix} package.}
\label{fig:measurement_plot} 
\end{figure*}

We calculate integrated measurements from the SOFIA polarization and APEX CO data within 90 hexagonal apertures representing individual sight-lines into the galaxy. The hexagonal apertures tessellate the galaxy and are spaced at half the beam size of the APEX and SOFIA data ($\theta_{\rm APEX}/2 = 10''$). We choose this aperture size as a compromise between having sufficient samples in each region and avoiding correlated information in adjacent apertures (for further details on the calculations of measurements and their uncertainties, refer to Sect. \ref{sec:methods_hexaprops}). We plot the measurements in Figure \ref{fig:measurement_plot} and report their corresponding correlation coefficients. We label points in these plots according to regions in the galaxy as in LR23, which divides the galaxies into the overlap region, the tidal tails, the relic (northern) spiral arm, and the galaxy cores (See Sect. \ref{sec:methods_regassign} for details of the region masking). The corresponding assignment of the hexagonal apertures is shown in the optical image panel of Fig. \ref{fig:presentation}.



We first examine global trends in Figure \ref{fig:measurement_plot}. The top-left plot reflects the anti-correlation in polarization fraction ($p$) vs CO intensity ($I_{\rm CO}$) seen in the maps, with the scatter in $p$ decreasing as a function of $I_{\rm CO}$. The polarized intensity ($PI$) on the other hand, appears to be independent of $I_{\rm CO}$ (middle-left panel). The tightest correlation appears between the FIR intensity ($I_{\rm FIR}$) and $I_{\rm CO}$ (bottom-left panel), a trend also apparent from their respective maps, since this plot shows the resolved Kennicutt-Schmidt relation. Resolving these relations region-wise, we notice a pattern across the plots involving $p$ where the spiral arm and tidal tail tend to have similar slopes (with the tidal tail exhibiting higher polarization fractions), while the overlap region shows a different slope. This is particularly apparent in the $p$ vs $I_{\rm CO}$ and $I_{\rm FIR}$ vs $I_{\rm CO}$ plots. The difference in behavior in the $p$ vs $I_{\rm CO}$ plot is possibly due to qualitatively different depolarization effects occurring in the overlap region -- which is more turbulent and has presumably more tangled B-fields as a result -- versus the spiral arms and tidal tails. In the $I_{\rm FIR}$ vs $I_{\rm CO}$ plot, the flatter slope for the overlap region is likely due to lower optical depths due to macroscopic velocities in the region. These effects are discussed in more detail in Sect. \ref{sec:disc_measurements}.

A similar global anti-correlation as seen between $p$ and $I_{\rm CO}$ is also present in the $p$ vs velocity dispersion ($\sigma_{\rm CO}$) plot (top-middle panel), but with a consistent scatter vs $\sigma_{\rm CO}$. This is in part a consequence of $\sigma_{\rm CO}$ at kpc-scales being non-trivial to interpret in a merger, since the variations in LOS velocity traced include contributions from kinematically distinct regions, such as between the southern galaxy and the overlap region. With the contribution of different regions highlighted, we see once again that the tidal tail and spirals arms are qualitatively similar, with the tidal tails appearing to just be offset with a higher $p$. Qualitatively however, the trends across regions are similar, with negative slopes for $p$, flat trends for $PI$, and positive slopes for $I_{\rm FIR}$ vs $\sigma_{\rm CO}$. We also plot $p$ against the dust temperature $T_{\rm dust}$ (top-right panel) and column density $N_{\rm H{_{2}}}$ (middle-right panel) as derived in LR23, and find globally negative trends with both quantities. In these trends, the tidal tail is quite disjoint from the rest of the galaxy, and has a notably steeper negative dependence on $T_{\rm dust}$, however we do not make strong inferences from this given the small range in $T_{\rm dust}$ values (25-29 K).


The anti-correlations seen in $p$ are in contrast to the flat trends displayed by $PI$, which appears to be independent of $\sigma_{\rm CO}$ and $I_{\rm CO}$, with the exception of the overlap region showing a modest positive slope ($b\sim0.2$) vs $I_{\rm CO}$. LR23 found a stronger positive trend holds for $PI$ vs $N_{\rm HI-H_{2}}$, particularly in the overlap region ($b\gtrsim0.56$). We interpret this as a consequence of the $I_{\rm CO}$ not accurately tracing the column density, which is an indicator of optical depth effects in the region.

To summarize the key results of the comparison of the APEX and SOFIA data: Globally, the polarization fraction anti-correlates with the CO surface density and velocity dispersion, as well as with the dust temperature and column density. The overlap region showing a weaker anti-correlation with the surface density relative to the relic spiral arm and tidal tail.The polarized intensity on the other hand is independent within uncertainties of either molecular gas property. The tight correlation between $I_{\rm FIR}$ vs $I_{\rm CO}$ shows the spatially-resolved Kennicutt-Schmidt relation.

\subsection{Comparison of ALMA structures and SOFIA B-fields}\label{sec:res_ALMA_SOFIA_comparison}

\begin{figure}
\centering
\includegraphics[width=0.5\textwidth]{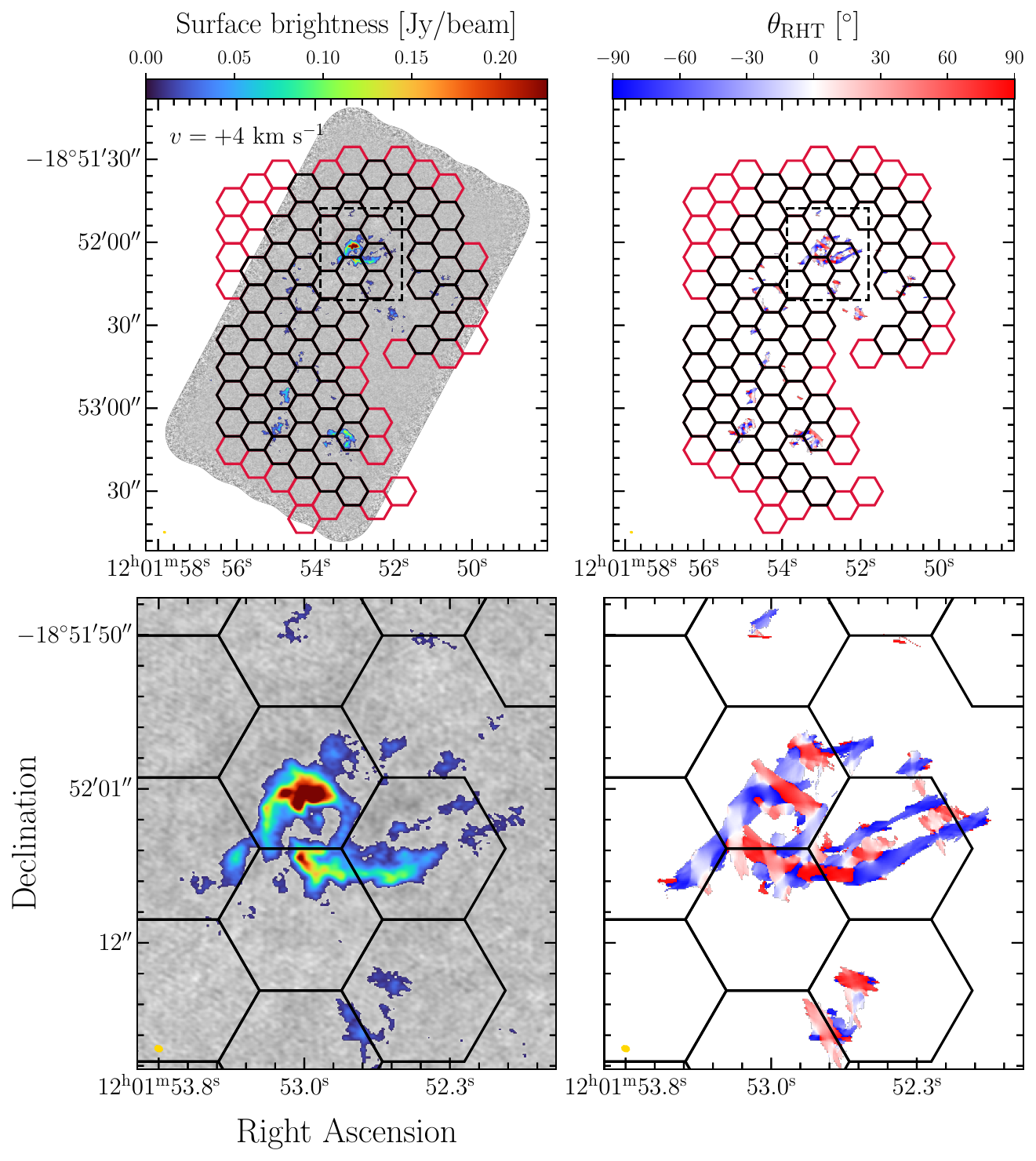}
\caption{Example application of the Rolling Hough Transform on the $v = +4$ km s$^{-1}$ channel. (\textit{Left column}) ALMA CO(3-2) intensity emission in the selected velocity channel, after masking. (\textit{Right column}) RHT-derived angles for the same channel. The top row shows the intensity and angles for the whole galaxy, while the bottom row zooms in on the core of the northern galaxy NGC 4038. The RHT is applied on the masked CO channel to avoid spurious features. The red hexagons mark apertures that are included for the comparison between APEX and SOFIA data, but are excluded for the ALMA analysis due to no overlap with the ALMA integrated intensity map.}
\label{fig:rht_example}
\end{figure}

The comparison of the APEX and SOFIA measurements highlight depolarization effects in the galaxy, specifically the $p$ vs $I_{\rm CO}$ and $p$ vs $\sigma_{\rm CO}$ plots. In a complex gas configuration such as the Antennae, we expect a case where B-field lines overlap from numerous regions along the LOS and are increasing tangled due to the injected turbulence from intense star formation. Since measurements of polarization signatures are a density-weighted vector sum of fields within the resolution element, more complex fields average and lower the measured degree of polarization, are thereby the product of effects below the beam size.  

The high resolution ALMA data allows us to peer within individual sight lines to observe the gas distribution on scales smaller than the APEX and SOFIA beams. Through the methodology followed in Sect. \ref{sec:methods_rht} (using the Rolling Hough Transform, or RHT), we quantify the pixel-by-pixel orientation of the CO emission depending on the surrounding linear structure they belong to, on a channel-by-channel basis. We then perform the following tests: (a) quantify the alignment of molecular gas with the measured B-field per aperture, and locate the velocities at which the maximum alignment occurs, and (b) quantify the geometric complexity of CO structures along the line-of-sight using the derived orientations within each aperture, akin to the methodology followed in \cite{2018ApJ...857L..10C}, and compare this with the measured SOFIA polarization fraction.

These tests are driven by the following rationale: For (a), we expect that a relatively denser gas component along the LOS would dominantly contribute its B-field orientation to the orientation measured by SOFIA. Locating a component along the LOS that is highly aligned with the continuum-derived B-field would therefore suggest the presence of such an effect. A strong assumption here is that the geometry of molecular gas and B-fields are well-coupled at the $\sim$50 pc scales of ALMA. There are indications of such an alignment broadly existing in the Galaxy and in molecular clouds at GMC-scales, particularly at densities traced by low-$J$ CO lines \citep{2011Natur.479..499L, 2019ApJ...878..110F, 2023MNRAS.521.3830M, 2025ApJ...991..156P, 2026arXiv260611323B}. However, even disregarding this assumption, if apertures in a particular region of the galaxy align strongest in kinematically coherent structures -- as inferred from the velocities at which the alignment occurs -- this would establish the presence of coherent gas components that preferentially align with the measured B-field direction, holding implications about the physical nature of the components. In the case of (b), we aim to obtain a quantitative measure of geometric depolarization, assuming geometrically complex gas harbors similarly complex B-field configurations. This has been shown in \cite{2018ApJ...857L..10C} with a comparison of HI fibers and \textit{Planck}-derived polarization fractions in the Milky Way, for which an HI ``coherence'' was calculated. Here we perform the same calculation to derive a CO coherence and make a similar comparison for an extragalactic source.

As an example of the CO orientations derived for the ALMA channels, we plot the results of the RHT applied on the systemic velocity channel in Figure \ref{fig:rht_example}. Within each channel, we obtain the alignment of gas structures with the measured B-field per-aperture using an alignment parameter $\zeta$ described in Sect. \ref{sec:methods_zeta}, such that $\zeta =1$ $(=-1)$ in the case of strong (poor) alignment. We therefore have the variation in the alignment of molecular gas structures and the local B-field across the galaxy as a function of velocity. For these tests we restrict our analysis to apertures that overlap with the ALMA integrated intensity map, which reduces the number of apertures from 90 to 66. The difference in grids is shown in Fig. \ref{fig:rht_example}. 

\subsubsection{Alignment of line-of-sight components with the B-field in the overlap}\label{sec:res_ALIGNED_IN_OVERLAP}

\begin{figure*}[ht]
\centering
\includegraphics[width=1\textwidth]{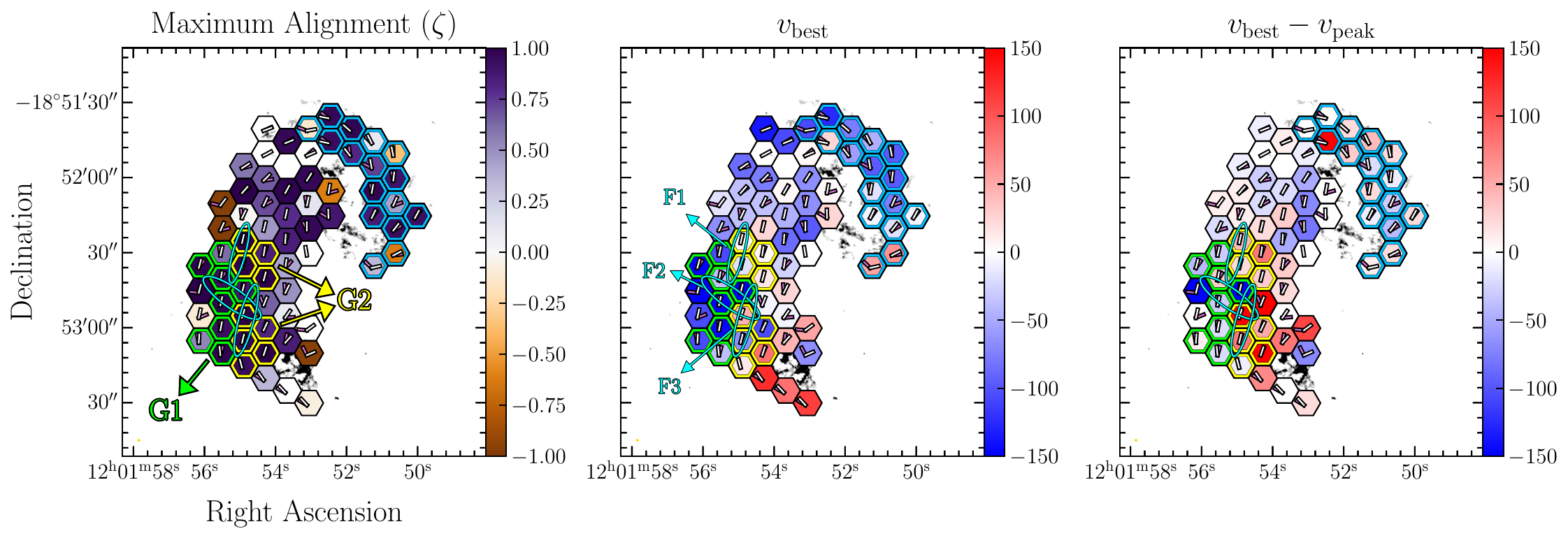}
\caption{Maps showing for each aperture, (\textit{Left}) the alignment between CO emission and B-fields ($\zeta$) obtained at $v_{\rm best}$, (\textit{Middle}) the velocity of best alignment, and (\textit{Right}) the best alignment velocity with respect to the peak-intensity velocity of the aperture-integrated CO spectrum. In the plots, the white vectors show the directions of the dust polarization vectors rotated by 90$^{\circ}$ to show the B-field direction, while the purple vectors show the intensity-weighted orientation of the gas, at the best aligned channel for each aperture. The hexagons outlined in green (G1) and yellow (G2) show groups harboring similar $v_{\rm best}$s, but which are kinematically distinct from each other, i.e. the G1 is blue-shifted w.r.t. the G2. The cyan ellipses mark the positions of filamentary structures identified in ALMA CO(3--2) data by \cite{2014ApJ...795..156W}.}
\label{fig:rht_2comps}
\end{figure*}

In order to look for velocity components strongly aligned with the measured B-field, we obtained the variation of $\zeta$ as a function of velocity for each aperture. We then performed a simple rolling average of $\zeta$ w.r.t the velocity with a window size of 3 channels to avoid spikes due to stochastically high alignments, and identified peaks in the resulting $\zeta$ variation and the velocities at which they occur. The obtained velocities of best alignment ($v_{\rm best}$) were then compared with the velocity at which the integrated CO spectrum of the apertures peak ($v_{\rm peak}$), as a proxy of the density of components along the LOS.

The outcome of this test is shown in Figure \ref{fig:rht_2comps}, which shows the maximum alignment (i.e. $\zeta$ computed at $v_{\rm best}$), $v_{\rm best}$, and $v_{\rm best}-v_{\rm peak}$. This gives us a first indication if there are specific structures lending their intrinsic field orientations to the observed B-fields. In the northern spiral arm, we find that $v_{\rm best}$ varies smoothly in velocity and is blue-shifted as a whole, showing that the apertures belong to a coherent velocity component. This is unsurprising given sight-lines in the relic spiral arm are kinematically simple and primarily comprise of a single component. The fact that we also see a reasonably strong gas/B-field alignment (Fig. \ref{fig:rht_2comps}, first panel) shows that the spiral arm component is aligned with the B-field, confirming that the spiral B-field vectors seen in the SOFIA data also follow gas structures down to the $\sim50$ pc structures seen by ALMA. Interestingly, even in the spectrally complex overlap region, we find two distinct groups of apertures that appear coherent, which we highlight in the $v_{\rm best}$ and $v_{\rm best}-v_{\rm peak}$ plots. Specifically, the green group (G1) of apertures point towards a component at $\sim$0 km s$^{-1}$, while the yellow group (G2) arises from a blue-shifted component at $\sim-166$ km s$^{-1}$. The groups are also strongly aligned with the B-field, with $\zeta$ values of 0.94 and 0.79, respectively. The $v_{\rm best}-v_{\rm peak}$ plot shows that the two groups broadly that the two components and the spiral arm occur at velocities close to velocities where the spectra in the apertures peak, i.e. they are close in velocity to the brightest spectral components, with median $v_{\rm best}-v_{\rm peak}$ values of -25 km s$^{-1}$, 50 km s$^{-1}$, and 15 km s$^{-1}$ for the G1, G2, and spiral group of apertures, respectively. If the $v_{\rm peak}$ is taken to be a proxy of the velocity of the densest component along the LOS, this is indicative that the densest component attributes its B-field orientation on the large-scale B-field. In Fig. \ref{fig:rht_2comps}, we also highlight the locations of filamentary structures identified by \cite{2014ApJ...795..156W} using CO(3--2) observations. The spatial locations of F1 and F3 correspond closely with the northern and southern parts of G2, and also appear at similar velocities ($\bar{v}_{\rm F1, F3}\sim-40$ km s$^{-1}$, $v_{\rm G2, \zeta_{peak}}\sim 0$ km s$^{-1}$). This is also the case for F2 which lies along G1, and shares the same kinematics ($\bar{v}_{\rm F2}\sim-180$ km s$^{-1}$, $v_{\rm G1, \zeta_{peak}}\sim -166$ km s$^{-1}$). This shows that the identified groups are clearly associated with previously identified filamentary molecular gas structures, and that the latter are preferentially aligned with the ordered B-fields of the galaxy.

\begin{figure*}
\centering
\includegraphics[width=1\textwidth]{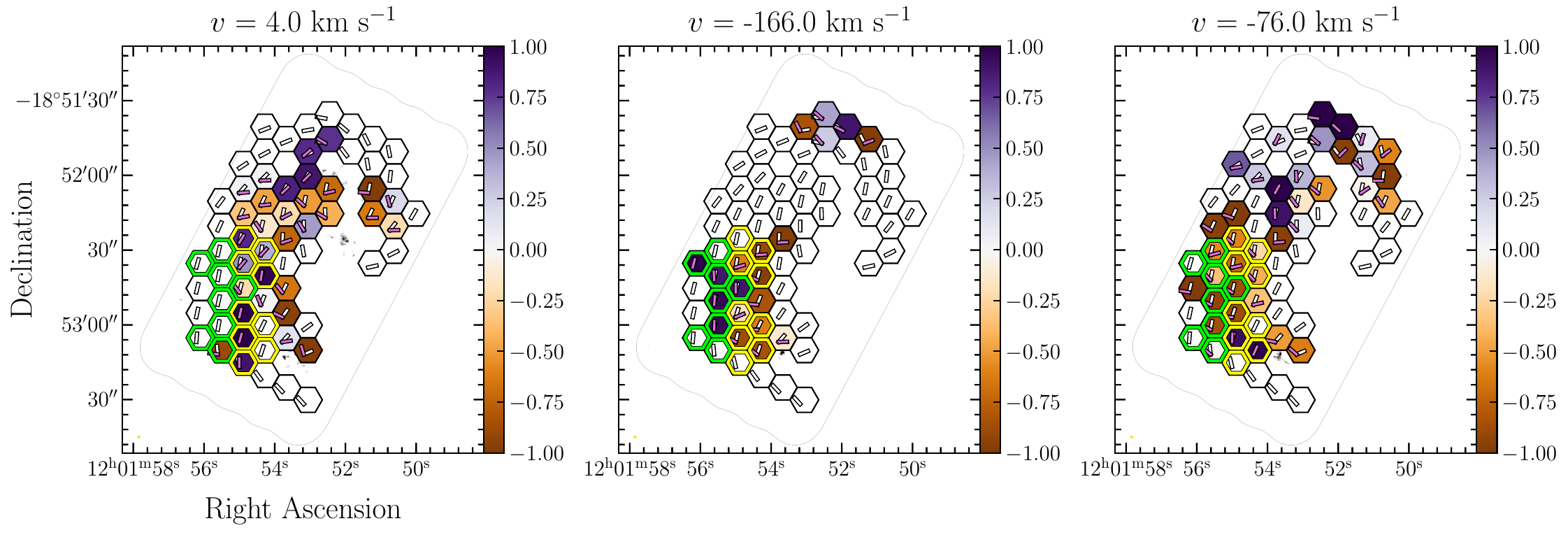}
\caption{Alignment of groups in the overlap region as a function of velocity. Each plot shows the degree of alignment $\zeta$ as filled colors, and shows the SOFIA/HAWC+ polarization vectors rotated by 90$^{\circ}$ to show the B-field direction in white, and intensity-weighted orientation in each aperture for the channel. The green (G1) and yellow (G2) groups are kinematically distinct groups identified in the overlap region, which is shown in this figure. At $v=+4$ km s$^{-1}$, G1 shows excellent alignment, while G2 does not yet contain significant emission. At $v=-166$ km s$^{-1}$ G1 is poorly aligned with the molecular gas, while G2 now aligns with the gas. At an intermediate velocity $v=-76$ km s$^{-1}$ we see neither group being well aligned with the gas structures at the associated velocity, showing that the alignment does not trivially improve with the presence of CO emission.}
\label{fig:rht_compvels}
\end{figure*}

To check if the presence of the alignment of the groups at these velocities is simply coincident with the fact that there is emission at those velocities, we plot in Figure \ref{fig:rht_compvels} the alignment $\zeta$ at the velocities of the identified groups, and a third distinct velocity where sufficient emission occurs at the spatial positions of the groups. The identified groups as a whole are poorly aligned with the B-field at velocities away from their best alignment velocities, at which the groups show strong overall alignments ($\bar{\zeta}_{\rm G1} > 0.94$, $\bar{\zeta}_{\rm G2} > 0.79$). These results therefore show that we do in fact observe strongly aligned components along the LOS with coherent velocities, over $\sim$5 apertures ($\sim5$ kpc). As mentioned in the previous section, there are physical implications for this result, in that these components could be dominant contributors to the measured field direction owing to their densities. Alternatively, it is possible that these structures host higher degrees of polarization relative to other spectral components. We discuss these interpretations further in Sect. \ref{sec:disc_components}.

\subsubsection{A CO coherence map of the Antennae}\label{sec:res_COHERENCE_MAP}

\begin{figure*}
\centering
\includegraphics[width=1\textwidth]{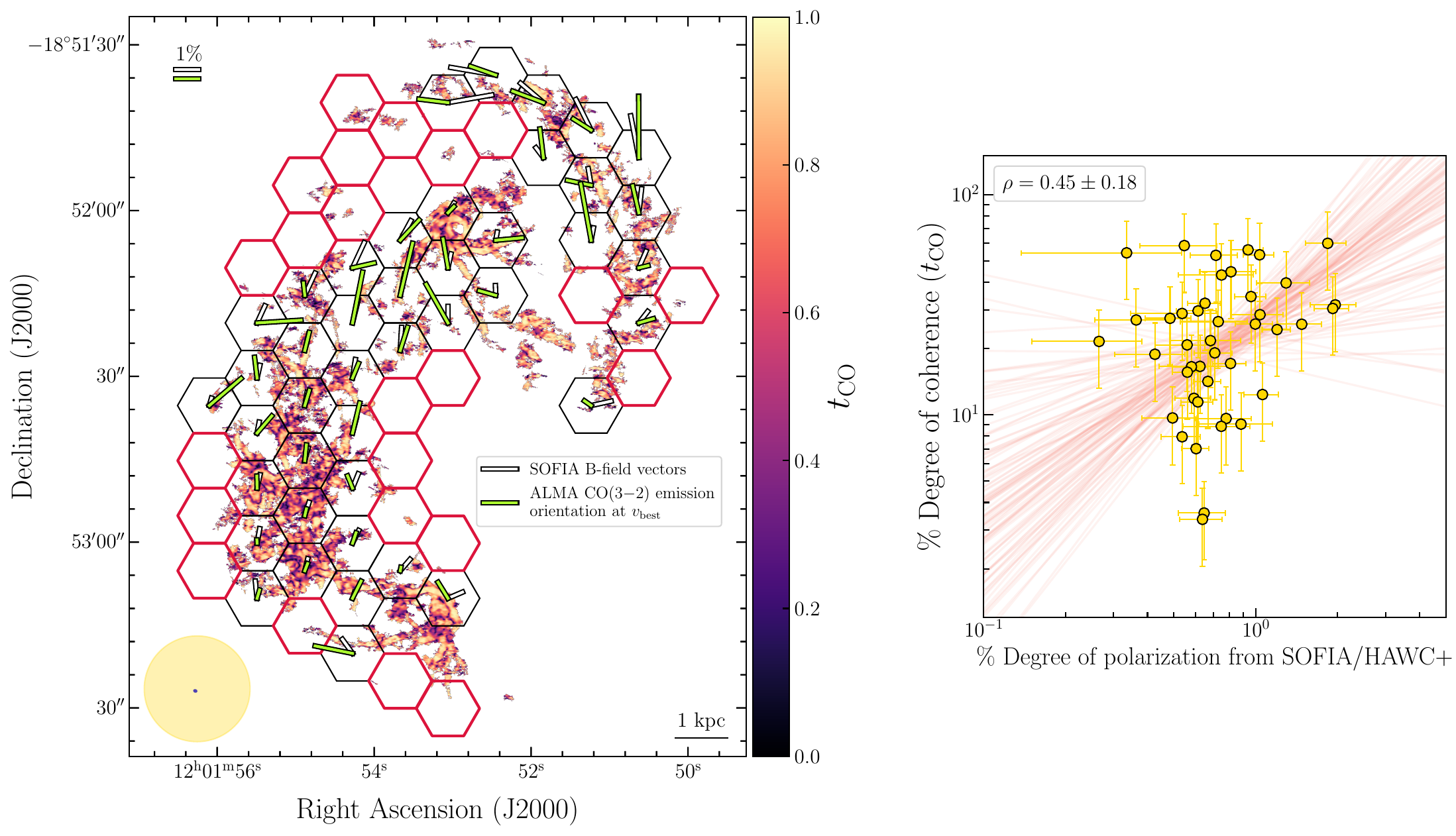}
\caption{Coherence of CO structures, calculated for apertures with a filling fraction of atleast 15\% (excluded marked in red). (\textit{Left}) Map of $t_{\rm CO}$ measuring angular variance along the line-of-sight obtained from the RHT analysis. The white vectors show proportional to the degree of polarization in the inferred direction of the B-field, while the purple vectors show the intensity-weighted orientation of CO structures at the $v_{\rm best}$ of the apertures. Both quantities are calculated at the scale of the apertures. The beams of the SOFIA (gold) and ALMA (blue) data are shown in the bottom left, highlighting the difference in scales. (\textit{Right}) Scatter plot of the measured $t_{\rm CO}$ and degree of polarization $p$ with uncertainties. The correlation coefficient $\rho$ was computed using the \texttt{linmix} package. The salmon lines show samples from the posterior distribution of the linear fit.}
\label{fig:rht_pcoh}
\end{figure*}

To further trace a link between the polarization data from SOFIA and the CO data from ALMA, we compute a map of the coherence of CO emission in the galaxy using the orientation of CO structures, computed using the RHT technique on a channel-by-channel basis (for further details, refer to Sect. \ref{sec:methods_rht}). The coherence, $t_{\rm CO}$, is a metric of the angular dispersion along the spectral axis, defined so that for large variations in angle, $t_{\rm CO}$ is close to 0, while for small variations, $t_{\rm CO}$ is closer to 1. The resulting $t_{\rm CO}$ map shown in the left panel of Figure \ref{fig:rht_pcoh}. The lowest values across the galaxy are seen in the overlap region, particularly within the interiors of the giant molecular clouds, and close to the cores of both galaxies. The higher values of $t_{\rm CO}$ appear in filamentary structures across the entire galaxy, close to the northern core and in the spiral arms of both galaxies. This is a trend also broadly seen in the degree of polarization of the SOFIA data.

For comparison with the previously calculated $p_{\rm hex}$ (in Sect. \ref{sec:res_APEX_SOFIA_props}) We compute an intensity-weighted coherence for each aperture. Given the relatively sparse ALMA emission we ensure that the apertures contain emission in at least 15\% of their area. The resulting comparison is the right panel of Fig. \ref{fig:rht_pcoh}. The correlation found is modest, with $\rho\sim0.5$. A stronger correlation ($\rho\sim0.7$) is seen between MW HI fibers and polarization fractions from \textit{Planck} \citep{2018ApJ...857L..10C} in the Milky Way. We note a couple of caveats here: The insensitivity of the high resolution ALMA data to diffuse and extended structures in the tidal tails means we do not sample this correlation at points where $p$ is the highest. Additionally, unlike the analysis in \cite{2018ApJ...857L..10C} we do not smooth $Q_{\rm CO}$ and $U_{\rm CO}$ to the resolution of the polarization observations. The reasoning for this is two-fold: since the polarization is inherently a function of structures finer than the beam size, we posit that using the un-smoothed coherence is necessary to infer the presence of a link between the distinct scales of the SOFIA and ALMA data. Additionally, the large contrast ($\sim$40 times) in the resolution of ALMA and SOFIA means that smoothing to the SOFIA beam would significantly dampen the imprint of any signal in the ALMA map. This is a primary reason for the HI fibers appearing less coherent with a median $t_{\rm HI}\sim0.15$. This effect is precisely what we observe when using a smoothed coherence, as demonstrated in Figure \ref{fig:smoothing}.

\section{Discussion}\label{sec:disc}

\subsection{Insights from APEX and SOFIA comparison}\label{sec:disc_measurements}

A common trend we see in Fig. \ref{fig:measurement_plot} is the decrease of the polarization fraction with increasing intensity, velocity dispersion, column density, and temperature, whereas the polarized intensity remains relatively constant with the same quantities. The presence of such depolarization indicates a variety of underlying physical mechanisms \citep{2015A&A...576A.104P}. Geometric depolarization can occur due to the effect of frequent reversals in the B-field orientation along the LOS, due to tangling of B-field lines within the observational beam, or due to different field lines lying along the same LOS. Depolarization could also be a consequence of weakening of grain alignment mechanisms due to increasing temperature, or in dense environments which can shield grains from the local UV radiation field which helps spin up the grains to supra-thermal energies which drives the alignment with the local B-field \citep{1915PhRv....6..239B, 1976Ap&SS..43..291D}.

In the case of the Antennae galaxies, an ongoing merger with high kinematic complexity and high turbulence, we expect the effects of the geometric depolarization to dominate. In fact, even in the case of the more quiescent spiral galaxy M51, the alignment efficiency of grains with the local B-field was found to dominate over depolarization effects due to collisional damping, as well as alternative alignment mechanisms with the local UV radiation field or by mechanical alignment \citep{2024ApJ...975..119L}.

With this in mind, the $p-I_{\rm CO}$ relation at higher intensities probes denser regions where we expect depolarization to be stronger, and thus expect a negative slope. We find that the slope of the $p-I_{\rm CO}$ plot in the spiral arm and tidal tails is similar, whereas the overlap region has a flatter $p-I_{\rm CO}$ slope, indicating depolarization towards gas rich regions is less effective. This is surprising given the overlap region is most affected by the dynamics of the merger, hosts intense star formation that injects turbulence in the ISM, and is thus likely to host a complex B-field configuration more susceptible to depolarization. Some insight into understanding this apparent discrepancy is offered by optical depth effects, wherein the large velocity differences in the overlap region reduce the optical depth and lead to a smaller $\alpha_{\rm CO}$ \citep{2024ApJ...971..176H,  2003ApJ...588..243Z, 2023A&A...673A..13M}, which means the high CO intensity regions ($\gtrsim15$ Jy arcsec$^{-2}$) in the overlap are not much denser regions as they are chaotic. This also explains a similar anomaly in the $I_{\rm FIR}-I_{\rm CO}$ plot, where the slope -- a proxy for the star forming efficiency -- is lower for the overlap region. The fact that $p$ shows little decline as a function of $I_{\rm CO}$ despite the increasing turbulence then implies a B-field with a higher ordered-to-random field ratio that is insensitive to the depolarization, perhaps due to stretching caused by the merger. We note here that a way to circumvent the optical depth issue is to instead look at the $p-N_{\rm HI-H_{2}}$ relation, with $N_{\rm HI-H_{2}}$ derived from optically thin FIR emission, which is shown in LR23, however while this brings the slope of the overlap region more in line with the rest of the galaxy, the overlap still has a broadly similar slope to the relatively undisturbed spiral arm, and highly anisotropic tidal tail, which means the interpretations discussed above might still be at play.



The variation of $p$ with $\sigma_{\rm CO}$ more closely probes the depolarization mechanisms involved, since as noted previously $\sigma_{\rm CO}$ not only traces intrinsic gas turbulence but also includes bulk cloud motions in the beam area as well as along the line-of-sight. LR23 find from a visual inspection of SOFIA data (same as used in this work) and smoothed ALMA CO(1--0) data, that $PI$ appears to anti-correlate with $\sigma_{\rm CO}$ over the entire galaxy, with the latter being lower in the tidal tails and highest in the overlap. While we do see a broad anti-correlation between individual $PI$ and $\sigma_{\rm CO}$ extrema points, with high and low $PI/\sigma_{\rm CO}$ values in the tidal tail and overlap, respectively, the individual trends within each region yield relatively flat relations. For the individual points, the differences in $PI$ can be seen as an increase in anisotropic turbulence due to compression and shearing of the gas in the tails, while the corresponding increase in the overlap and decrease in $PI$ is explained by stronger isotropic random fields driven by a turbulent dynamo mechanism due to feedback from star formation and supernovae in the overlap. We thus expect the slope of $p-\sigma_{\rm CO}$ to be steepest in the overlap, instead we find similar slopes in $p-\sigma_{\rm CO}$ across all the regions. This tells us that the effects of depolarization are broadly similar despite differences in the anisotropy of the turbulence and gas configuration. It is worth noting that the northern galaxy also contains a dispersion feature tracking the northern spiral arm in the APEX dispersion map, which appears to trace a kinematic offset of the northern spiral relative to the northern galaxy and extends to the base of the tidal tails, and such bulk features could be responsible for causing similar depolarization effects in relatively undisturbed regions. The fact that we obtain flat regional trends for $PI$ vs $\sigma_{\rm CO}$ could also reflect this behavior. 

\subsection{Implications for high-redshift polarimetric observations}\label{sec:highz_implications}

The moderate resolution of the SOFIA and APEX data is similar to the kpc-scale physical resolutions probed by ALMA at high redshift \citep{2023Natur.621..483G,2024A&A...692A..34C}. The proximity of the Antennae merger however enables us to supplement our analysis with the resolved structure of the galaxy at hand. Spectrally, the resolved data helps pinpoint kinematic structures associated with the polarized emission with more accuracy. \cite{2024A&A...692A..34C} studied the ordered field in the lensed galaxy SPT0346-52 at $z\sim5.6$, with a physical resolution of 700 pc at the region of highest amplification. They performed a rough comparison of the polarized intensity map with [\textsc{C ii}] emission, and identified a single gas velocity component coinciding with the location of the polarized emission. In the case of the Antennae we more specifically identify sub-regions within the galaxy where specific velocity components coincide and coherently align with the magnetic field direction. By locating these components through our analysis of the alignment of the FIR and high resolution CO emission, our study shows that the unresolved coincidence of gas emission and polarization can be pinned down to specific gas structures producing the measured polarization signature.

The high resolution information also helps alleviate degenerate interpretations. From the moment maps in Fig. \ref{fig:presentation} it is clear that at these kpc-scales no region is necessarily devoid of macroscopic velocity gradients, which can complicate interpretations of variations of polarized quantities with the total line intensity and the velocity dispersion. We also note that the velocity fields and dispersion maps themselves can produce features that can be misleading, such as the interface of the overlap and southern galaxy which closely resembles disk rotation, and the corresponding high dispersion feature which can be interpreted as a minor axis outflow. In the absence of complementary high resolution data, interpretations of high redshift polarization data with kpc-scale spectroscopic data thus warrants caution.




\subsection{Source of dominantly aligned components along the line-of-sight}\label{sec:disc_components}


The results from Sect. \ref{sec:res_ALMA_SOFIA_comparison} highlight distinct kpc-scale regions in the overlap along the LOS that preferentially align with the density-weighted average polarization direction inferred by SOFIA. The fact that we see aligned molecular gas components in the overlap is interesting in itself, since the overlap consists of intense star formation and is the main interaction region of the merger, where given sufficient turbulence it could be naively expected that B-fields would be sufficiently randomized so that any memory of the alignment of molecular gas structure with the polarization direction is lost. The presence of an ordered B-field in spite of such an effect suggests that this is not the case, meaning the region still hosts an ordered B-field. An ordered B-field in such a chaotic region could be the sign of compressive shearing motions in the overlap, possibly due to infalling motions through the effect of gas losing angular momentum during the collision, leading to an large-scale anistropic field. A scenario where the contribution of the merger can be neglected could also be true; \cite{2004A&A...417..541C} roughly estimate that the magnetic diffusion timescale is of the order of the dynamical timescale of the merger estimated from simulations, and thus the remnant of the ordered B-fields once present in the galaxy could still survive in the overlap region.

The broad alignment observed between the B-field orientation and the geometry of the low density phase of the molecular gas at clouds scales enables us to obtain some insight on the aligned components seen in the overlap as being contributors to the field orientation seen at kpc-scales. The propensity of a component along that LOS that would dominantly lend their orientation onto the kpc-scale B-field measurement is inherently a function of the intrinsic polarization and density of the structure. The two regions contain potentially show a key difference in this regard. While most regions within G1 and G2 belong to the brighter spectral component along the LOS (as identified by $v_{\rm best}-v_{\rm peak}$ values closer to 0 in Fig. \ref{fig:rht_2comps}), both groups contains apertures that belong to the fainter spectral component. This naturally produces different interpretations: when the dominantly aligned component is denser, compressive gas motions could enhance the density of the molecular gas, alongside increasing B-field strengths and the strength of the polarization signature. In the case of the less dense component aligning the mean polarization, the denser region presumably hosting more star formation could see its intrinsic polarization signature scrambled by the turbulence injected by stellar feedback. 

\cite{2004A&A...417..541C} from synchrotron polarization measurements use the fact that the increase in B-field strengths in the overlap are in accordance with an enhancement via compression ($B\propto\rho^{2/3}$) and that large magnetic diffusion timescales exist in the overlap region to attribute the observed B-fields to be more in line with compression being a dominant mode of B-field enhancement rather than a small-scale turbulent dynamo. This order of dominance might not necessarily follow down to the cold ISM, where the increased influence of stellar wind and supernova-driven turbulence might make a small-scale dynamo more relevant. Without a direct tracer of the density of the velocity components we observe, it is hard to endorse one mode of B-field enhancement over the other. However, we note that in the case turbulence tangles fields in the densest component leaving secondary components to determine the resulting field direction, we would not expect a coherent field direction over several kpc -- which is what we observe in the western overlap region component (highlighted in yellow in Figs. \ref{fig:rht_2comps} and \ref{fig:rht_compvels}). This gives us some indication that compressive flows might be responsible for B-field enhancement in the overlap, even in the cold ISM.

\subsection{Coherence of structures along the LOS}\label{sec:disc_coherence}

Our application of the RHT in the Antennae is adopted from the methodology of \cite{2018ApJ...857L..10C}, who perform a similar analysis on \textit{Planck} polarization and HI maps. A key result in their work is the connection between Galactic B-fields and HI structures, which is established from the correlation between the coherence of HI fibers and the \textit{Planck} degree of polarization. In a similar vein, Fig. \ref{fig:rht_pcoh} gives us an idea of the robustness of such a coupling in the case of the Antennae. Specifically, the ``coupling'' here would probe the correlation between the complexity of molecular gas structures, and the complexity of B-field lines.

A key difference between our work and \cite{2018ApJ...857L..10C} are the gas structures under consideration. Their study focuses on the coupling of high Galactic latitude diffuse HI fibers on pc-scales, which are very different from the kpc-scale molecular features studied here, and thus the coupling probed here shows that the effects of geometric depolarization are seen even at galaxy-scales. In addition, the presence of such a correlation shows that the polarization measured at the beam size of SOFIA traces the complexity of the configuration of gas within the beam, which we resolve with the ALMA observations at scales of 50 pc. Using synthetic FIR polarimetric observations of galaxies, \citet{MA2024} found that observations with resolutions of $\sim300$ pc trace B-field configurations at the highest scales of the simulation at $\sim20$ pc. An interesting question that arises here is how the correlation of the coherence and polarization fraction measurements obtained at the ALMA resolution would appear. At the kpc scales traced by SOFIA, we expect only ordered fields to contribute to the measurement of $p$, since the disordered component of the field cancels out. At pc scales however, depending on the relative contribution of the disordered field component, one would expect the $p_{\rm CO}$ correlation to weaken if the contribution is strong, or remain the same or strengthen if not, and use this to set a scale at which disordered fields become relevant across different environments in the galaxy. One avenue to probe this is through cloud-scale full polarization observations of the dust continuum or of molecular spectral lines in the Antennae, which could also help specifically test the assumptions made in this paper of the correspondence of B-fields and molecular gas structures. In the Milky Way, such a study could be potentially achieved by the Atacama Large Aperture Submillimeter Telescope (AtLAST)\footnote{\href{www.atlast.uio.no}{www.atlast.uio.no}} \citep{2025A&A...694A.142M, 2026arXiv260705022C}, which would allow for observations of both spectral line polarization and continuum polarization with a matched observational setup. AtLAST would extend such work to the most diffuse and extended structures thanks to its extremely high mapping speed of $\sim2.5\times10^{3}$ deg$^{2}$ mJy$^{-2}$ hr$^{-1}$, reaching a regime which is currently prohibitively expensive to perform with ALMA. Additionally, AtLAST would be able to probe high frequency lines such as [CI](1--0) at 492 GHz which can help avoid the optical depth ambiguities posed by low-$J$ CO lines.

\section{Conclusion}{\label{sec:conclusion}}

We studied the link between molecular gas and magnetic fields in the Antennae galaxy merger, using new APEX CO(3--2) observations, complemented by archival SOFIA/HAWC+ dust polarization and ALMA CO(3--2) observations. We first compared the resolution-matched new APEX and archival SOFIA data to quantitatively infer how polarization in the galaxy varies as a function of the kinematics and geometry of the bulk gas. The ALMA observations then allowed us probe below the APEX and SOFIA beam sizes and quantify the geometry of the pc-scale molecular gas configuration as a function of velocity. Leveraging the fact that the continuum polarization is a density-weighted average along the line-of-sight, we use the data to obtain tomographic insights into magnetic fields in the overlap region. Our key findings are summarized below:
\begin{enumerate}
    \item The comparison of the APEX and SOFIA data show strong depolarization with both CO intensity and velocity dispersion ($b\lesssim -0.4$), which we interpret as strong depolarization owing to the presence of high turbulence and overlapping B-field structures in the overlap region, and compressive/shearing processes enhancing polarization in the tidal tails.
    \item The overlap region shows weaker depolarization compared to the tidal tail and the relic northern spiral arm in the galaxy. This is indicative of a strongly ordered B-field in the overlap region, possibly amplified by compressive flows.
    \item We locate coherent velocity components in the overlap region that are strongly aligned ($\zeta>0.79,0.94$) with the magnetic field direction from SOFIA relative to other components lying along the same sight-line. Assuming that the cloud-scale magnetic field follows the orientation of molecular gas structures at these scales, this implies that these components dominantly impress their B-field orientations onto the field direction measured at kiloparsec scales. This also shows that the broader polarization direction inferred by SOFIA can be traced down to specific gas structures located far below the beam size. Once again, we find that compressive enhancement of B-fields a likely scenario occurring in the overlap region.
    \item We use the orientation of gas structures as a function of velocity to compute a map of the coherence of CO structures. The coherence of the structures computed at the scale of the SOFIA dust polarization data shows a modest ($\rho\sim0.5$) with the degree of polarization, which reflects the correspondence of geometric complexity in the gas at pc-scales with the polarization fraction measured at kiloparsec scales.
\end{enumerate}


\begin{acknowledgements}
AL would like to thank the University of South Carolina, Columbia for their hospitality during the research stay, where part of this work was carried out. AL acknowledges financial support from a Kristine Bonnevie stipend from the University of Oslo. This project uses data collected under the Atacama Pathfinder EXperiment (APEX) Project (ID: 0112.F-9507), led by the Max Planck Institute for Radio Astronomy at the ESO La Silla Paranal Observatory. This paper makes use of the following ALMA data: ADS/JAO.ALMA\#2021.1.00439.S. ALMA is a partnership of ESO (representing its member states), NSF (USA) and NINS (Japan), together with NRC (Canada), NSTC and ASIAA (Taiwan), and KASI (Republic of Korea), in cooperation with the Republic of Chile. The Joint ALMA Observatory is operated by ESO, AUI/NRAO and NAOJ. This research has made use of the NASA/IPAC Extragalactic Database (NED), which is funded by the National Aeronautics and Space Administration and operated by the California Institute of Technology. Based on observations made with the NASA/DLR Stratospheric Observatory for Infrared Astronomy (SOFIA) under the 08\_0012 Program. SOFIA is jointly operated by the Universities Space Research Association, Inc. (USRA), under NASA contract NNA17BF53C, and the Deutsches SOFIA Institut (DSI) under DLR contract 50 OK 0901 to the University of Stuttgart. This project has received funding from the European Union's Horizon Europe research and innovation programme under grant agreement No. 101188037 (AtLAST2). Views and opinions expressed are however those of the author(s) only and do not necessarily reflect those of the European Union or European Research Executive Agency. Neither the European Union nor the European Research Executive Agency can be held responsible for them.

\textit{Software:} \textit{AstroPy} \citep{2013ASPC..475..241T}, \textit{Matplotlib} \citep{Hunter:2007}, \textit{SciPy} \citep{2020SciPy-NMeth}
\end{acknowledgements}

\bibliographystyle{aa}
\bibliography{bibliography}

\begin{appendix}

\section{Methods}\label{sec:methods}

We describe here the methods to enable the comparison of molecular line data from ALMA and APEX with the polarization data from SOFIA. We overlay a hexagonal grid over the galaxy, representing sight lines within which we calculate physical quantities across the datasets. The apertures are spaced at the Nyquist sampling limit, at half the APEX beam size ($\theta_{\rm APEX}/2 = 10''$). The tiling is shown in Figure \ref{fig:presentation}.

\subsection{Calculation of molecular gas and polarization quantities}\label{sec:methods_hexaprops}

With the goal of a resolved comparison of the molecular gas and B-field properties across the merger, we calculate properties of both datasets on an aperture-wise basis. From the CO data, we compute in each hexagonal aperture the integrated CO intensity and CO velocity dispersion, and their corresponding errors. From the SOFIA data, we use the Stokes $I$ map to calculate the integrated FIR intensity, and the median Stokes $Q$ and $U$ in an aperture to recompute the polarized intensity and polarization fraction per aperture. That is,
\begin{equation}\label{eq:p_hex}
    p_{\rm hex} = \frac{\sqrt{\langle Q\rangle^{2} + \langle U\rangle^{2}}}{\langle I\rangle},
\end{equation}
where $\langle I\rangle$, $\langle Q\rangle$, $\langle U\rangle$ represent median Stokes $I, Q, U$ values within an aperture. The corresponding error in the polarization fraction is:
\begin{equation}
    \sigma_{p} = \frac{\sqrt{\sigma_{Q}^{2} + p^{2}\sigma_{I}^{2}}}{\langle I\rangle},
\end{equation}
where $\sigma_{Q}$ and $\sigma_{I}$ are median uncertainties within the apertures calculated from the error maps from the SOFIA data.

The polarization fraction is subsequently de-biased to eliminate noise-induced polarization fractions:
\begin{equation}\label{eq:p_hex_deb}
    p_{\rm hex,deb} = \sqrt{p_{\rm hex}-\sigma_{p}^{2}}.
\end{equation}
From this, we obtain the de-biased polarized intensity as:
\begin{equation}\label{eq:pi_hex}
    PI_{\rm hex,deb} = p_{\rm hex,deb} \langle I\rangle.
\end{equation}

We also used column density and dust temperature maps derived in LR23 using the SOFIA and archival \textit{Herschel} PACS and SPIRE data and computed the mean H$_{2}$ column density and dust temperature in the hexagonal apertures.

To quantify relationships between the pairs of quantities, we model a power-law relation, which we fit for by plotting them on a log-log scale and performing a linear regression:
\begin{equation}\label{eq:power_law}
    \log{y} = a + b\log{x}
\end{equation}

For the latter, we use the Bayesian linear regression method by \cite{2007ApJ...665.1489K}, \texttt{linmix\_err}, which accounts for uncertainties in both dependent and independent variables, and outputs a posterior distribution for the fit parameters, as well as the strength of the correlation between two quantities. For our purposes, we use the median and the standard deviation of the posterior distribution as a measure of the best-fit and 1$\sigma$ uncertainty on the relations.



\subsection{Orientation of B-fields and ALMA molecular gas structures}\label{sec:methods_rht}

The obtained B-field orientations from dust polarization is a density-weighted vector average of B-field orientations within the resolution element of the observation \citep{2015ASSL..407...59A,MA2024}. The high resolution ALMA data (0.52$''$/55 pc) which resolves features $\sim36$ times smaller than the SOFIA beam size (19$''$/2 kpc), enables for a direct comparison between B-field and molecular gas orientations within the apertures.

The intensity-weighted orientation of the B-field in an aperture can be calculated from the median Stokes $Q$ and $U$ parameters of the SOFIA data,
\begin{equation}\label{eq:pa_hex}
    \theta_{\rm hex} = \frac{1}{2} \arctan\Big(\frac{\langle U\rangle}{\langle Q\rangle}\Big).
\end{equation}
which is related to the orientation of the magnetic field by,
\begin{equation}\label{eq:pa_to_b_hex}
    \theta_{\rm B} = \theta_{\rm hex} + \frac{\pi}{2}.
\end{equation}

For the ALMA observations, we use the Rolling Hough Transform (RHT), described in detail in \citep{2014ApJ...789...82C}, developed to quantify linear structure and the orientation of HI fibers in the Milky Way. Briefly, the first step of the process is performing an unsharp mask by subtracting the original image by a smoothed version of the image (using a kernel of diameter \textit{D$_{K}$}), thus suppressing large-scale structure in the image, after which the image is thresholded at 0 to create a bitmask. This ensures the detection of linear feature is agnostic to the brightness of the features. Next, a disk of diameter \textit{D$_{W}$} is extracted from each point in the image which represents the domain over which a Hough transform is performed. While the original Hough transform is typically applied so that every point $(x,y)$ is mapped to every possible line in $(\rho,\theta)$ space (where $\rho = x\cos{\theta}+y\sin{\theta}$), the RHT is applies the Hough transform keeping $\rho=0$. The outcome is a one-dimensional spectrum of linear intensity $R(x, y, \theta)$ as a function of $\theta$ which encodes the relative contribution of each angle $\theta$ to the structure at $(x,y)$. Finally, only pixels above a linear power threshold $Z$ are retained.

For our analysis, we re-bin the masked ALMA data cube to a velocity resolution of 10 km s$^{-1}$, and apply the RHT method on 60 velocity channels that contain CO(3--2) emission, specifically ranging from -367 km s$^{-1}$ to 223 km s$^{-1}$, which helps us identify linear structures in the molecular gas as a function of velocity, in a process similar to that described in \cite{2018ApJ...857L..10C}. We tested a variety of parameter values of ($D_{K}$, $D_{W}$, $Z$) and chose parameters of (1.51$''$, 3.44$''$, 0.5) or (18 pixels, 41 pixels, 0.5) for the RHT on each channel. The estimation of the orientations of these structures can be obtained as following. The linear intensity $R(x, y, \theta)$ per velocity channel can be decomposed into its Stokes \textit{Q}$_{\rm RHT}$ and \textit{U}$_{\rm RHT}$ components, as:
\begin{align}
    & Q_{\rm RHT}(x, y, v) = \int \cos{(2\theta)} \, R(x,y,\theta, v) d\theta \label{eq:Q_RHT} \\
    & U_{\rm RHT}(x, y, v) = \int \sin{(2\theta)} \, R(x,y,\theta, v) d\theta.\label{eq:U_RHT}
\end{align}
These can be used to estimate the orientation at each ($x, y$):
\begin{equation}\label{theta_RHT}
    \theta_{\rm RHT} = \frac{1}{2} \arctan\Big(\frac{U_{\rm RHT}}{Q_{\rm RHT}}\Big).
\end{equation}

The average molecular gas orientation $\theta_{\rm CO}$ for an aperture is then computed similar to Eq. \ref{eq:pa_hex}, using the median of $Q_{\rm RHT}$ and $U_{\rm RHT}$.

Having obtained the orientations at each point as a function of velocity, we can extract a degree of coherence, similar to that defined in \cite{2018ApJ...857L..10C} for HI fibers, for the linear CO structures in the velocity channels. $t_{\rm CO}$, which we term the `CO coherence', is analogous to the polarization fraction $p$, and is bounded between [0, 1]. This is computed through analogous Stokes $Q$ and $U$ parameters using the RHT-derived angles and the CO intensity from the ALMA CO(3--2) cube:
\begin{align}
    & Q_{\rm CO}(x, y) = \int I_{\nu} \cos{(2\theta_{\rm RHT})} \, dv, \label{eq:Q_CO}\\
    & U_{\rm CO}(x, y) = \int I_{\nu} \sin{(2\theta_{\rm RHT})} \, dv, \label{eq:U_CO}
\end{align}
where $I_{\nu}$ is the CO intensity at ($x, y$) at the velocity channel $v$. The $t_{\rm CO}$ parameter is then calculated as:
\begin{equation}\label{eq:t_co}
    t_{\rm CO} = \frac{\sqrt{Q_{\rm CO}^{2}+U_{\rm CO}^{2}}}{I_{\rm CO}},
\end{equation}
where $I_{\rm CO}$ is the integrated CO intensity over the same range of velocity channels for which the RHT is computed. The $t_{\rm CO}$ value for an aperture is computed similarly, using the median $Q_{\rm CO}$ and $U_{\rm CO}$ values in an aperture. The CO coherence as defined here measures the variation in the orientation of structures throughout this range in velocity. If the orientation changes greatly in this range, $t_{\rm CO}$ will be low, whereas a relatively constant angle will result in a $t_{\rm CO}$ close to 1.

\subsection{Quantifying the alignment of B-fields and molecular gas}\label{sec:methods_zeta}

Now that we have the orientation of the B-field and molecular gas available for each aperture, we can define their mutual alignment, $\zeta(\theta_{\rm B}, \theta_{\rm CO})$ similar to that defined in \cite{2019ApJ...887..136C}. We first define the difference between the angles, $\delta\theta$ as,
\begin{equation}\label{eq:deltheta}
    \delta\theta = \frac{1}{2} \arctan\left[\frac{\sin{(2\theta_{\rm CO})}\cos{(2\theta_{\rm B})} - \cos{(2\theta_{\rm CO})}\sin{(2\theta_{\rm B})}}{\cos{(2\theta_{\rm CO})}\cos{(2\theta_{\rm B})} + \sin{(2\theta_{\rm CO})}\sin{(2\theta_{\rm B})}}\right].
\end{equation}
$\delta\theta$ is then used to compute the alignment parameter $\zeta$, as,
\begin{equation}\label{eq:zeta}
    \zeta(\theta_{\rm CO}, \theta_{\rm B}) = \cos(2\delta\theta).
\end{equation}
$\zeta$ defined as such has a range from [-1,1], wherein if the orientations are perpendicular, $\zeta = -1$, and for parallel alignments, $\zeta = 1$.

\subsection{Region-wise assignment of hexagonal apertures}\label{sec:methods_regassign}

We assign each aperture in the hexagonal grid to a region of the galaxy as identified in LR23, who use the \texttt{astrodendro} program \citep{Rosolowsky_2008} to split the galaxy into 5 regions, namely the tidal tails, the relic northern spiral arm, the overlap region, and the cores of the NGC 4038 and NGC 4039. We match their mask, shown in LR23 (their Fig. 5) to the pixel grid of the data used in this work, and assign each aperture to the corresponding region in which the center of the aperture lies.

\section{Spectral profile of APEX dispersion feature}

\begin{figure*}
\centering
\includegraphics[width=0.85\textwidth]{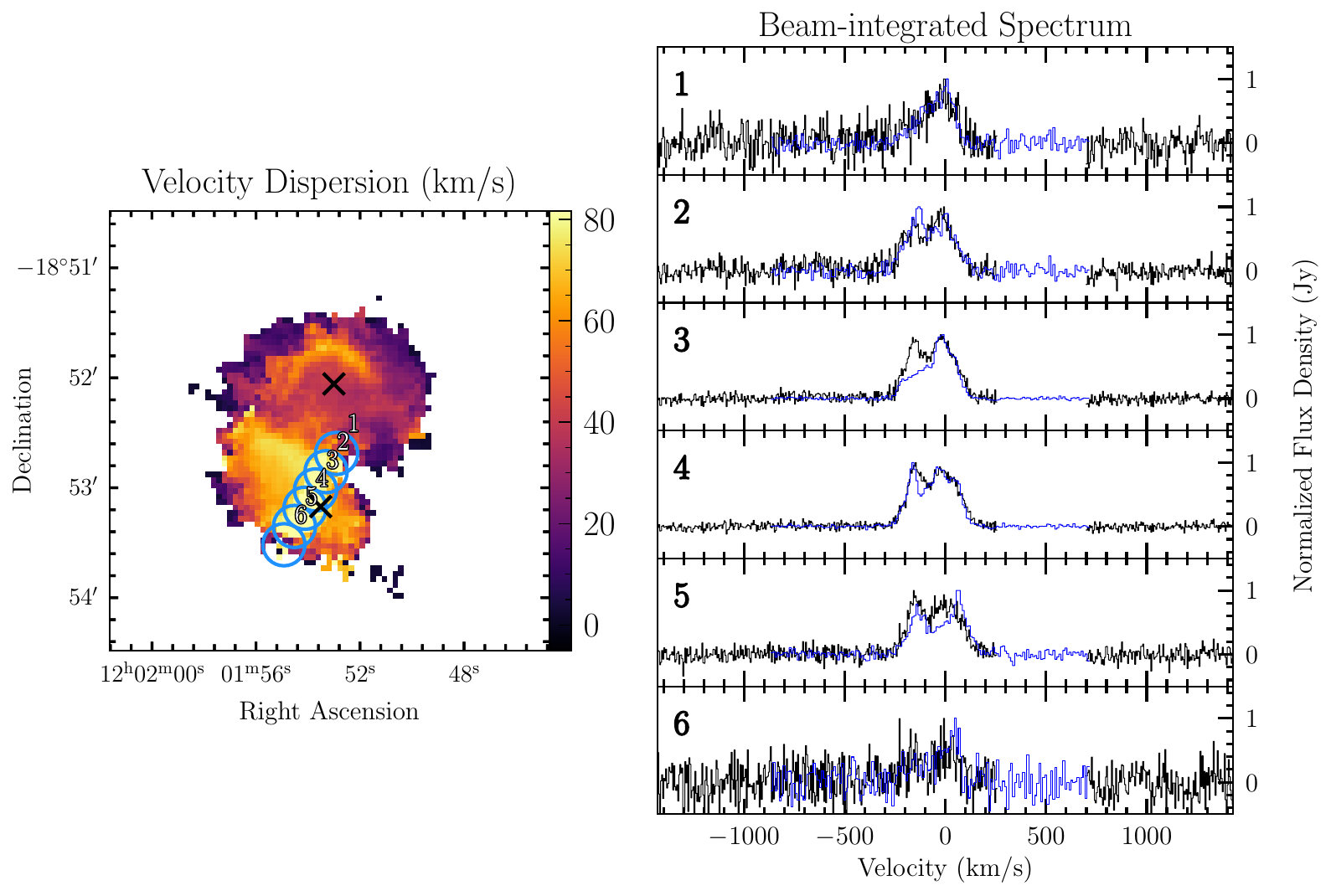}
\caption{No clear evidence of outflows in the high $\sigma_{\rm CO}$ feature. (\textit{Left}) APEX moment-2 map showing the beam-sized apertures as blue circles along the high velocity dispersion feature in which corresponding spectra are calculated. The X's mark the centers of both galaxies. (\textit{Right}) Peak-normalized integrated spectra of the apertures from APEX (black) and ALMA (blue) CO(3--2) data. The gap in the APEX spectra are channels contaminated with an atmospheric ozone emission line, which we masked in the data reduction.}
\label{fig:highdisp_spec}
\end{figure*}

We compute integrated spectra from the APEX and ALMA data in apertures lying along the high velocity dispersion feature seen in the APEX moment-2 map, and show them in Figure \ref{fig:highdisp_spec}. While there is some broadening seen in aperture 3, the presence of outflowing motions is not readily evidenced from the APEX and ALMA spectra, and the increased dispersion is more simply explained by the two distinct velocity components from the overlap region and southern galaxy being contained within a single beam, which is most apparent in the spectra from apertures 2, 3, 4, and 5.

\section{Effect of smoothing on CO coherence}

\begin{figure*}
\centering
\includegraphics[width=1\textwidth]{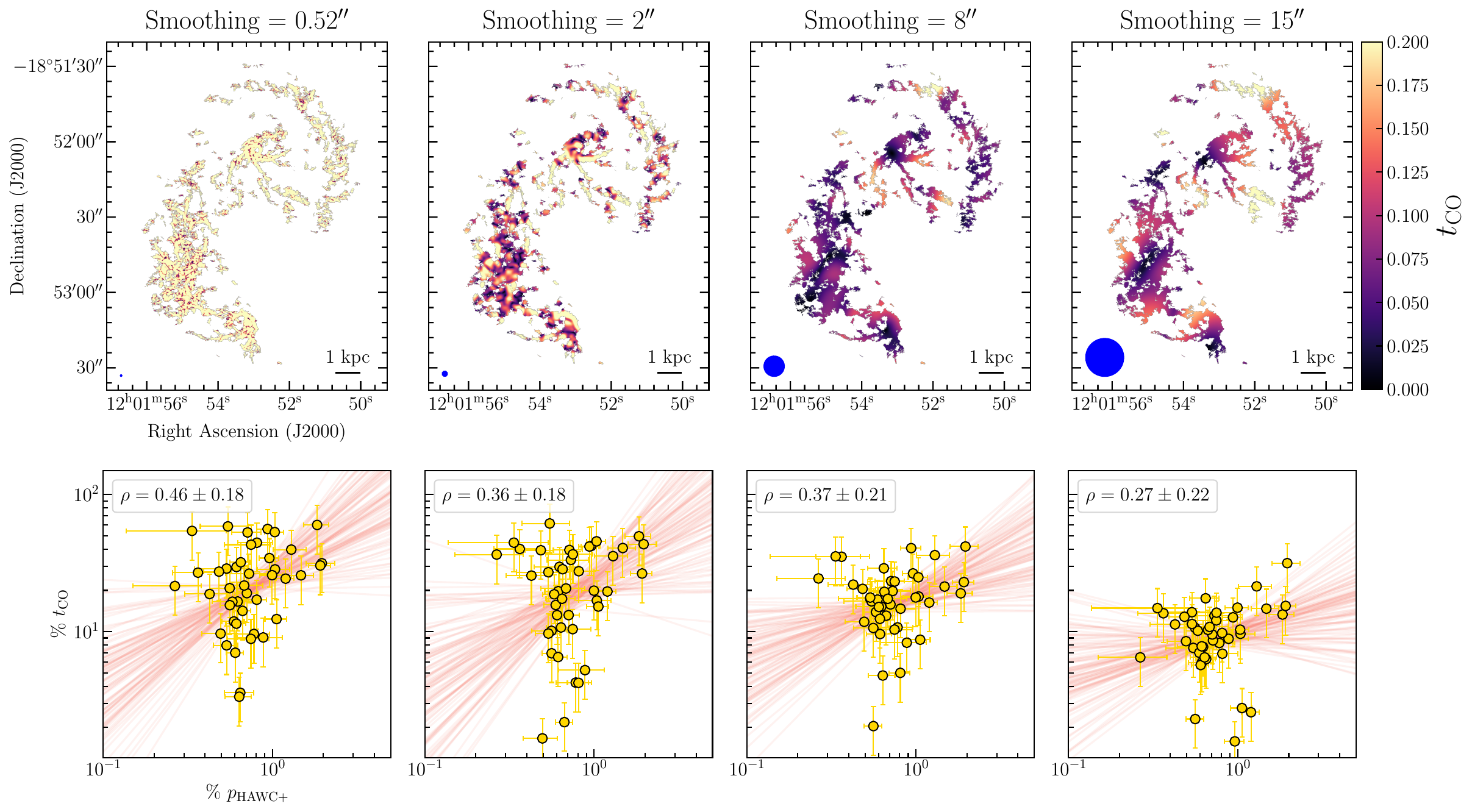}
\caption{Effect of smoothing on coherence maps and $t_{\rm CO}$-$p_{\rm hex}$ correlation. (\textit{Top row}) Coherence maps at 0.5$''$, 2$''$, 8$''$, 15$''$, calculated from smoothed $U_{\rm CO}$ and $Q_{\rm CO}$ maps. The colormaps of each plot are scaled to the same $t_{\rm CO}$ values for visualization. Beam sizes of each map are shown in blue in the bottom left. (\textit{Bottom row}) Correlation of $t_{\rm CO}$ and $p_{\rm hex}$ at different resolutions. $t_{\rm CO}$ values were calculated for the same hexagonal aperture grid used throughout the work using $U_{\rm CO}$ and $Q_{\rm CO}$ values in each aperture. The salmon lines sample the posterior distribution of the fit correlation coefficient, using \texttt{linmix\_err}.} 
\label{fig:smoothing}
\end{figure*}

We show here the results of calculating the correlation between $t_{\rm CO}$ and $p$, upon smoothing the $t_{\rm CO}$ (specifically, smoothing $U_{\rm CO}$ and $Q_{\rm CO}$ from which $t_{\rm CO}$ is computed). We smooth $U_{\rm CO}$, $Q_{\rm CO}$ to resolutions of 0.5$''$ (native), 2$''$, 8$''$, and 15$''$ to calculate $t_{\rm CO}$ and perform a similar linear regression versus $p_{\rm hex}$ with \texttt{linmix\_err} as in Sect. \ref{sec:methods_rht}, and present the results in Figure \ref{fig:smoothing}.

We find from the maps the larger smoothing values lose details of features in the overlap and spiral arm, with the largest smoothing at 15$''$ (still higher resolution than the 19$''$ SOFIA map) only retaining the lower coherence in the overlap, and high coherence in parts of the spiral arm. The trend in the correlation coefficients with larger smoothing is the decrease of the correlation strengths from 0.46 at 0.5$''$ to 0.27 at 15$''$, with an increase in the correlation uncertainty. The degree of coherence also decreases, with a median coherence of 24\% at 0.5$''$ to 10\% at 15$''$. These results highlight the necessity of using the ALMA data at native resolution for this comparison, where $p_{\rm hex}$ is intrinsically a function of features below the 19$''$ beam size, which are traced by the ALMA data.

\end{appendix}

\end{document}